\documentclass[11pt]{article}
\usepackage[sort&compress,square,comma,numbers]{natbib}

\usepackage[labelfont=bf]{caption} 
\usepackage[font={small}]{caption}   

\usepackage{subcaption}
\usepackage{amsmath}
\usepackage{amssymb}
\usepackage{url}
\usepackage{placeins}
\usepackage{textcomp}
\usepackage{graphicx}
\usepackage{float} 

\usepackage[usenames,dvipsnames]{color}    

\usepackage{soul}   
\setstcolor{red} 
\setul{}{1.5pt}  

\newcommand{\pd}[2]{\frac{\partial #1}{\partial #2}} 
\title{\textbf{Brillouin Cooling in a Linear Waveguide}}
\author{Yin-Chung Chen, Seunghwi Kim, Gaurav Bahl$^{\ast}$\\
	\\
	\footnotesize{Mechanical Science and Engineering, University of Illinois at Urbana-Champaign,}\\
	\footnotesize{Urbana, Illinois, USA}\\
	\footnotesize{$^\ast$To whom correspondence should be addressed; E-mail: bahl@illinois.edu}
	}

\date{}

\begin{document}
\sloppy

\maketitle

\begin{abstract}
Brillouin scattering is not usually considered as a mechanism that can cause cooling of a material due to the thermodynamic dominance of Stokes scattering in most practical systems. However, it has been shown in experiments on resonators that net phonon annihilation through anti-Stokes Brillouin scattering can be enabled by means of a suitable set of optical and acoustic states. The cooling of traveling phonons in a linear waveguide, on the other hand, could lead to the exciting future prospect of manipulating unidirectional phonon fluxes and even the nonreciprocal transport of quantum information via phonons. In this work, we present an analysis of the conditions under which Brillouin cooling of {phonons of both low and high group velocities }may be achieved in a linear waveguide. We analyze the three-wave mixing interaction between the optical and traveling acoustic modes that participate in forward Brillouin scattering, and reveal the key regimes of operation for the process. Our calculations indicate that measurable cooling may occur in a system having phonons with spatial loss rate that is of the same order as the spatial optical loss rate. If the Brillouin gain in such a waveguide reaches the order of 10$^{5}$ m$^{-1}$W$^{-1}$, appreciable cooling of phonon modes may be observed with modest pump power of {a few} mW.
\end{abstract}

\section{Introduction}

\begin{figure}[H]
	\begin{center}
		\includegraphics[trim=4.9cm 0.3cm 4.9cm 0.4cm, clip, width=0.75\textwidth]{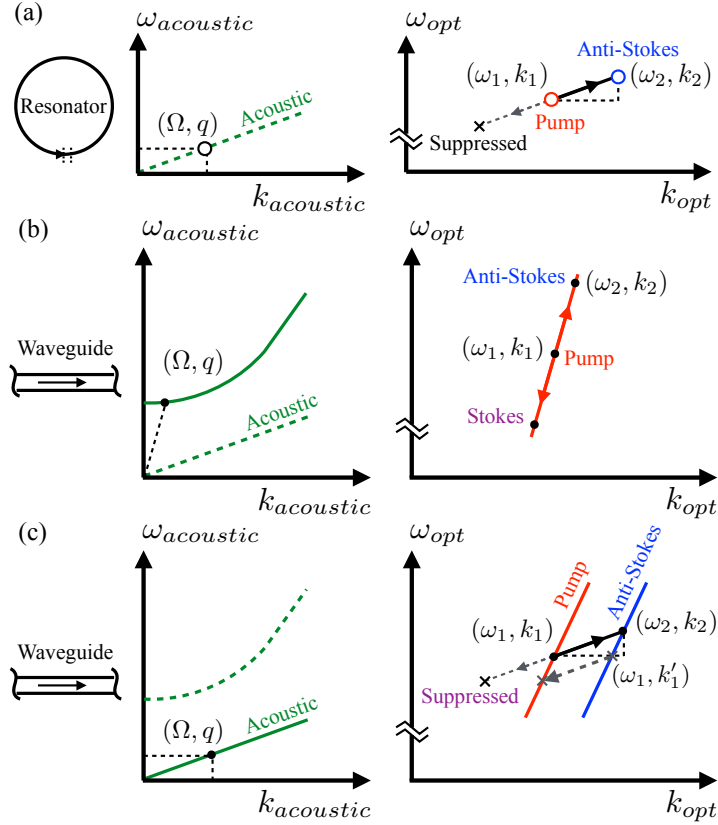}
		\caption{Phase-matching conditions for Brillouin cooling. 
			\textbf{(a)} In a resonator, the available modes are represented by discrete points (circles) in the $\omega-k$ space. Stokes scattering can be suppressed by the naturally available high-order modes of a resonator, and only the anti-Stokes photons are permitted, leading to net phonon annihilation. In a waveguide, due to lack of modal confinement in the direction of propagation, the discrete mode points become discrete mode lines. The scattering process can be categorized into two categories:
			\textbf{(b)} Intra-modal scattering. In this case, the phonon in the process belongs to low group velocity guided-acoustic mode. Since there exists at least one phonon mode with frequency and momentum ($\omega,q$) that matches the slope of the optical band, the Stokes process cannot be suppressed.
			\textbf{(c)} Inter-modal scattering. The phonon in the scattering process belongs to a mode family having large group velocity and appreciable momentum. If we excite the pump light only at the mode with momentum $k_1$, Stokes scattering is suppressed which potentially enables cooling if other required conditions are met. At the same time, we must avoid exciting the mode at $k_1'$ which will lead to Stokes scattering with the same phonon mode.			}
		\label{fig:PM}
	\end{center}
\end{figure}

Spontaneous Brillouin scattering of light from acoustic phonons \cite{shen1965theory, boyd2003nonlinear} results in creation of \textit{photons} at both higher (anti-Stokes) and lower (Stokes) energies than the pump. {Since} the scattering process is inelastic, additional \textit{phonons} are simultaneously created in the material during Stokes transitions, while existing phonons are annihilated in anti-Stokes transitions. Such annihilation of phonons from a mode can be broadly termed as a cooling phenomenon as it reduces the vibrational energy present in the solid \cite{arcizet2006radiation,gigan2006self,kleckner2006sub,bahl2012observation}.
In a bulk isotropic material, {however}, the probability of spontaneous Stokes scattering is always higher than that of anti-Stokes scattering \cite{peter2010fundamentals}. Furthermore, the possibility of achieving Stokes-directed \textit{stimulated} Brillouin Scattering (SBS) is quite high due to the very large Stokes gain. As a result, net phonon creation and heating of the system are nearly always encountered in experiments. 
The solution to the fundamental challenge of achieving net phonon annihilation lies in {engineering} the Stokes vs.\ anti-Stokes scattering probabilities through the photonic density of states of the {system} \cite{purcell1946spontaneous, gaponenko2002effects,chen2015raman}.
In practice, this can be achieved through a photonically-engineered structure like a resonator or photonic crystal \cite{gong2010linewidth, checoury2009enhanced, boroditsky1999spontaneous}. 
Exploitation of this idea previously led to the demonstration of Brillouin cooling in whispering-gallery resonators \cite{bahl2012observation}, where selection of discrete modes in energy-momentum space permitted the complete suppression of Stokes scattering (Fig.~\ref{fig:PM}a). 
Indeed, a second requirement revealed in studies of resonant Brillouin cooling \cite{agarwal2013multimode,tomes2011quantum} is that the photon lifetime within the system must be smaller than the phonon lifetime. In other words, photons must radiate and transfer energy out of the system without being reabsorbed or scattering back. Finally, higher opto-acoustic coupling, i.e. higher Brillouin gain \cite{dostart2015giant}, also helps improve the cooling rates.

In this paper we address the question of whether {Brillouin cooling of traveling phonons of arbitrary group velocity} can be achieved in a linear waveguide. This question is motivated by three reasons, which collectively point towards the possibility of cooling traveling phonons in such a system.
First, recent experimental developments on linear waveguides have begun to report extremely large Brillouin gains \cite{rakich2010tailoring,rakich2012giant,van2015interaction} suggesting that sufficient gain may be available for practical cooling experiments.
Second, light exits a local region of a waveguide considerably faster than phonons do, suggesting that the local, apparent phonon lifetime is considerably longer than the photons. {Addressing this question effectively requires an analysis of both high and low group velocity phonons, since these spatial relationships can result in different conclusions.} As we show later, the spatial coherence length is the determining parameter rather than temporal residence time in the linear waveguide case.
Finally, it is practical to engineer the optical modes of a linear waveguide to achieve Stokes suppression and anti-Stokes enhancement (Fig. \ref{fig:PM}c), similar to how cooling was achieved in resonators.

Previously, the cooling of phonons in linear waveguides has been briefly discussed in the regime where the phonons are more spatially coherent than the photons \cite{van2016unifying}. However, there has been no discussion of the resulting phonon spectrum, or a substantial discussion of the regime where phonons are less spatially coherent than the photons.
To fully evaluate the phonon spectrum, a careful investigation of the Langevin noise force that drives the phonons is essential. However, previous studies on noise-initiated Brillouin scattering \cite{boyd1990noise,kharel2016noise} have only been performed on low group velocity phonons in which case the propagation term can be neglected in the acoustic equation. This approximation is not suitable for traveling phonons that have large group velocity.
In this paper, we aim to fill this gap in knowledge by studying the Brillouin cooling of large group velocity traveling phonon modes inside a waveguide, with the inclusion of Langevin noise force and determination of the resulting phonon spectrum.
Our analysis here presents the conditions under which appreciable cooling could potentially be achieved and hopes to help direct further experimental studies on linear waveguides.

\section{Coupled wave equations for anti-Stokes Brillouin scattering}
\label{sec:3W}

A preliminary requirement for cooling is the suppression of the Stokes scattering process. In a resonator, this can be achieved \cite{bahl2012observation} {through inter-modal scattering between high-order optical modes, which is generally asymmetric} (Fig. \ref{fig:PM}a). However, the situation in a waveguide is drastically different. First, the acoustic modes that participate in Brillouin scattering can be categorized into two families, namely the guided-acoustic modes (Fig. \ref{fig:PM}b) and the traveling-acoustic modes (Fig. \ref{fig:PM}c), leading to intra-modal scattering and inter-modal scattering respectively \cite{kang2011reconfigurable}. For intra-modal scattering, the Stokes process cannot be suppressed since there is always a phonon mode ($\Omega,q$) with extremely low group velocity that matches the slope of the optical band. Scattering of this form is {sometimes }termed as Raman-like \cite{butsch2014cw} or Guided Acoustic Wave Brillouin Scattering (GAWBS) \cite{shelby1985guided}. For inter-modal scattering, {on the other hand}, if we pump at the lower order optical mode only the anti-Stokes process {can be engineered to satisfy} the phase-matching condition while the Stokes process is suppressed. {This process is also routinely encountered} \cite{koehler2016resolving}. We emphasize here that it is critical to match the pump $k$ vector to the lower order mode, otherwise competing Stokes and anti-Stokes processes may arise, as illustrated in Fig \ref{fig:PM}c.

\begin{figure}[t!]
\begin{center}
\includegraphics[trim=2cm 3.2cm 2.1cm 3.1cm, clip, width=0.85\textwidth]{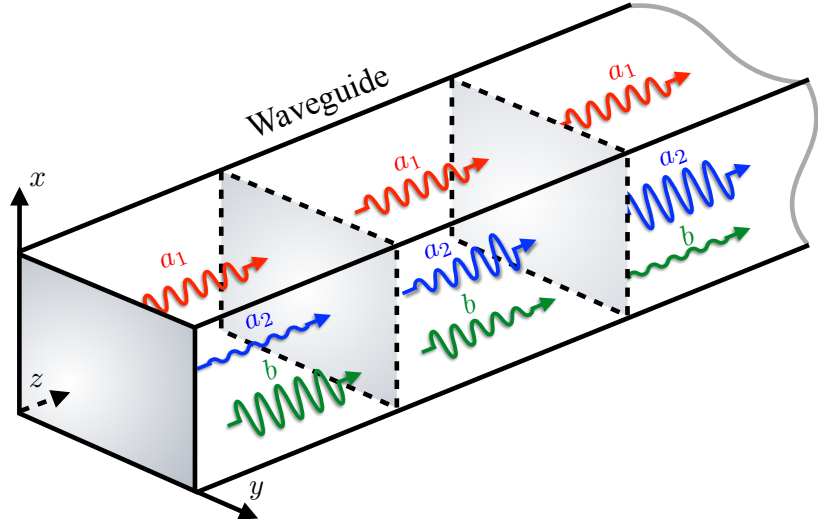}
\caption{Three wave mixing interaction in a linear waveguide. The waveguide is assumed to be infinitely extended in the $z$-direction. The pump field $(\omega_1, k_1)$, anti-Stokes field $(\omega_2, k_2)$, and the acoustic field $(\Omega, q)$ are described by their corresponding slowly varying field operators $a_1(z,t)$, $a_2(z,t)$, and $b(z,t)$, respectively.  We assume that the pump is non-depleted, i.e. $a_1(z,t)=\text{constant}$ throughout the waveguide, and the acoustic signal has a thermal equilibrium noise spectrum in the absence of coupling. As the three fields interact with each other inside the waveguide, the acoustic field experiences attenuation while the anti-Stokes field is amplified until they reach their respective steady state amplitudes.
		}
\label{fig:WG}
\end{center}
\end{figure}
%

Once process selection conditions are met, we can now consider the three-wave mixing process between a pump wave $(\omega_1, k_1)$, an anti-Stokes wave $(\omega_2, k_2)$, and an acoustic wave $(\Omega, q)$ in a linear waveguide. Energy and momentum conservation requirements set the relations ${\omega_1+\Omega=\omega_2}$ and ${k_1+q=k_2}$. As shown in Fig. \ref{fig:WG}, the three waves are assumed to co-propagate in a waveguide infinitely extended in the $z$-direction {for the forward scattering case}. {We consider here only the forward scattering case since the acoustic modes participating in backward scattering generally have greater frequency and thus lower $Q$-factor, which as we show later, are impractical to cool}. The coupling equations for the fields inside a waveguide can be derived classically as demonstrated in \cite{PhysRevA.92.013836} or quantum mechanically as in \cite{kharel2016noise,sipe2016hamiltonian}. Here we use the quantum description as it is more natural to introduce the Langevin noise operator that drives the acoustic field. Under the assumption that the process is perfectly phase matched, the equations of motion for the field operators are given by \cite{kharel2016noise,sipe2016hamiltonian}
\begin{subequations}\label{eq:3WQ}
	\begin{align}
	&\pd{a_1}{t}+v_1\pd{a_1}{z}
		=-\gamma_1a_1-i\beta^*a_2b^\dagger,
		\\
	&\pd{a_2}{t}+v_2\pd{a_2}{z}
		=-\gamma_2a_2-i\beta a_1b,
		\\
	&\pd{b}{t}+v_b\pd{b}{z}
		=-\Gamma b-i\beta^*a_1^\dagger a_2+\xi,
		\label{eq:3WQb}
	\end{align}
\end{subequations}
where the operators $a_1(z,t)$, $a_2(z,t)$, and $b(z,t)$ are the envelope operators of the pump, anti-Stokes, and acoustic field at their respective carrier frequencies $\omega_1$, $\omega_2$, and $\Omega$. Here $v_1$, $v_2$, and $v_b$ are the group velocities of the fields. It is important to note that with the definition of the field operators here, the expectation value of the form $\langle a^\dagger(z,t)a(z,t) \rangle$ should be understood as {photon (phonon) linear density (occupation number per unit length)} \cite{kharel2016noise}. The average power is therefore $P_1=\hbar\omega_1 v_1\langle a_1^\dagger a_1 \rangle$, $P_2=\hbar\omega_2 v_2\langle a_2^\dagger a_2 \rangle$, and $P_b=\hbar\Omega v_b\langle b^\dagger b \rangle$ for the pump, anti-Stokes, and acoustic field respectively. We have also introduced the temporal loss rates $\gamma_1$, $\gamma_2$, and $\Gamma$ of the fields and the Langevin noise \cite{boyd1990noise, agarwal2013multimode, kharel2016noise} operator $\xi$ into the coupled wave equations. Let us first discuss the behavior of the acoustic field with noise source when no coupling to the optical fields exists.

\subsection{Low group velocity acoustic modes at thermal equilibrium}
The properties of the acoustic noise source $\xi(z,t)$ in a waveguide have been derived previously in \cite{boyd1990noise,kharel2016noise} for low group velocity phonons. Here we briefly reproduce the derivation of \cite{kharel2016noise} as it will be important in our later discussion. To obtain the correlation relation of the noise operator, we first turn off the coupling term in the acoustic field equation. For a low group velocity phonon, the spatial variation of the acoustic field is negligible as it is small compared with the temporal variation term. We now divide the waveguide into small segments labeled by index $i$, each with length $\Delta z$ and uncorrelated noise source $\xi_i$. The equation satisfied by each $b_i(t)$ is
\begin{align}\label{eq:BI}
\pd{b_i}{t}=-\Gamma b_i+\xi_i.
\end{align}

The noise source of each segment obeys the following mean and correlation,
\begin{align}
&\langle \xi_i(t) \rangle=0,\\
&\langle \xi_i^{\dagger}(t)\xi_j(t') \rangle=Q\delta_{ij}\delta(t-t').
\end{align}
The factor $\delta_{ij}$ represents the fact that the noises in different segments are uncorrelated. From the solution to Eqn. (\ref{eq:BI}) we have
\begin{align}
b_i=\int_{-\infty}^{t}e^{-\Gamma(t-t')}\xi_i(t')dt',
\end{align}
and we can calculate the equal-time correlation
\begin{align}
\langle b_i^{\dagger}(t)b_j(t) \rangle
	&=\iint_{-\infty}^{t}\langle \xi_i^{\dagger}(t')\xi_j(t'') \rangle e^{-2\Gamma t}e^{\Gamma(t'+t'')}dt'dt'' \notag \\
	&=\int_{-\infty}^{t}Q\delta_{ij}e^{-2\Gamma t}e^{2\Gamma t'}dt'=\frac{Q\delta_{ij}}{2\Gamma}.
\end{align}
By definition, for $i=j$ this is also equal to the {phonon linear density}
\begin{align}
\langle b_i^{\dagger}(t)b_i(t) \rangle=\frac{n_{b,0}}{\Delta z},
\end{align}
where $n_{b,0}=[\exp(\hbar\Omega/k_BT)-1]^{-1}$ is the phonon occupation number of the acoustic mode at thermal equilibrium. By comparing the above two equations we find $Q=2\Gamma n_{b,0}/\Delta z$, and the correlation function becomes
\begin{align}\label{eq:NDis}
\langle \xi_i^{\dagger}(t)\xi_j(t') \rangle=\frac{2\Gamma n_{b,0}}{\Delta z}\delta_{ij}\delta(t-t').
\end{align}
Now we let $\Delta z \rightarrow 0$ and $i, j \rightarrow z, z'$ and $\delta_{ij}/\Delta z\rightarrow \delta(z-z')$, to obtain the mean and correlation for the noise operator in a continuous waveguide
\begin{align}
&\langle \xi(z,t) \rangle=0, \label{eq:NSM}\\
&\langle \xi^{\dagger}(z,t)\xi(z',t') \rangle=2\Gamma n_{b,0}\delta(z-z')\delta(t-t'). \label{eq:NS}
\end{align}

Here we note that these properties of the noise source lead to a divergent solution of the phonon density at a particular point in the waveguide $\langle b^{\dagger}(z,t)b(z,t) \rangle=\lim_{dz\rightarrow 0}\frac{n_{b,0}}{d z}$, which results from taking continuous limit of the discrete waveguide model. However, we can formally rewrite the phonon density as 
\begin{align}\label{eq:AK}
\langle b^{\dagger}(z,t)b(z,t) \rangle=\lim_{z\rightarrow 0}n_{b,0}\delta(z)=\lim_{z\rightarrow 0}\frac{1}{2\pi}\int_{-\infty}^{\infty}n_{b,0} e^{i kz} dk=\frac{1}{2\pi}\int_{-\infty}^{\infty}n_{b,0} dk.
\end{align}
The spectrum of the phonon density in the $k$-space is therefore flat and has a constant value $n_{b,0}$ {for this model of the noise source}. {In fact, as we will see later, important information about the phonon mode can be extracted from a narrow range in the $k$-space centered around $k=0$.}

\subsection{High group velocity acoustic modes at thermal equilibrium}
With the noise properties in Eqns. \ref{eq:NSM} and \ref{eq:NS}, we can now examine the behavior of high group velocity acoustic modes. The equation of motion for the acoustic field without coupling to the optical fields is given by
\begin{align}\label{eq:B}
\pd{b}{t}+v_b\pd{b}{z}=-\Gamma b+\xi.
\end{align}
To avoid the divergent problem we try to look at the power spectral density of $S_b(z,\omega)$ of $b(z,t)$. Defining the Fourier transform and the inverse transform by the expressions
\begin{align}
&\tilde{b}(z,\omega)=\int_{-\infty}^{\infty} b(z,t)e^{i\omega t}dt, \notag \\
&b(z,t)=\frac{1}{2\pi}\int_{-\infty}^{\infty} \tilde{b}(z,\omega)e^{-i\omega t}d\omega, \notag
\end{align}
The above Eqn. \ref{eq:B} can be written as
\begin{align}\label{eq:BFT}
-i\omega \tilde{b}+v_b\pd{\tilde{b}}{z}=-\Gamma \tilde{b}+\tilde{\xi}.
\end{align}
The solution for Eq. \ref{eq:BFT} is then obtained.
\begin{align}
\tilde{b}(z,\omega)=\tilde{b}(0,\omega)e^{-(\Gamma-i\omega)z/v_b}+\frac{1}{v_b}\int_{0}^{z}dz'e^{-(\Gamma-i\omega)(z-z')/v_b}\tilde{\xi}(z',\omega),
\end{align}
where $\tilde{b}(0,\omega)$ is the initial value of the field operator at $z=0$. Note that here we have also accounted for the propagation of the field and how the field $\tilde{b}(z,\omega)$ evolves under the influence of the noise source subject to the initial condition $\tilde{b}(0,\omega)$. 
The spectral density $S_b(z,\omega)$ of the acoustic field at position $z$ can be obtained by the relation \cite{agarwal2013multimode, mandel1995optical}
\begin{align}
\langle \tilde{b}^\dagger(z,\omega)\tilde{b}(z,\omega') \rangle=2\pi\delta(\omega-\omega') S_b(z,\omega).
\end{align}
It is reasonable to assume that the correlations $\langle \tilde{b}^\dagger(0,\omega)\tilde{\xi}(z',\omega') \rangle$ and $\langle \tilde{\xi}^\dagger(0,\omega)\tilde{b}(z',\omega') \rangle$ vanish, i.e. the acoustic field is uncorrelated to the noise at different spatial point. We can then invoke the properties of the noise to find
\begin{align}\label{eq:SbNC}
S_b(z,\omega)
	&=S_b(0,\omega)e^{-2\Gamma z/v_b}+\frac{2\Gamma n_{b,0}}{v_b^2}\int_{0}^{z}dz' e^{-2\Gamma (z-z')/v_b} \notag \\ 
	&=S_b(0,\omega)e^{-2\Gamma z/v_b}+\frac{n_{b,0}}{v_b}(1-e^{-2\Gamma z/v_b}),
\end{align}
where $S_b(0,\omega)$ is the spectral density of $b$ at $z=0$ under the initial condition $\tilde{b}(0,\omega)$. For $z\rightarrow\infty$ the spectral density is independent of its initial form $S_b(0,\omega)$ as anticipated.
\begin{align}\label{eq:SbNCInf}
S_b(\infty,\omega)=\frac{n_{b,0}}{v_b}.
\end{align}
From Eqn. \ref{eq:SbNC} we see that regardless of the detail of initial acoustic spectrum, the effect of the noise always bring the acoustic spectrum back to the value $n_{b,0}/v_b$. In fact, if we substitute Eqn. \ref{eq:SbNCInf} into Eqn. \ref{eq:SbNC} as the initial spectral density $S_b(0,\omega)$, we immediately find that the spectral density $S_b(z,\omega)$ is independent of position $z$ in the waveguide, as it should be in the case without coupling. 

The {phonon linear density} can be evaluated by integrating the spectral density over all frequencies:
\begin{align}\label{eq:AF}
\langle b^\dagger(z,t)b(z,t) \rangle=\frac{1}{2\pi}\int_{-\infty}^{\infty} S_b(z,\omega)d\omega=\frac{1}{2\pi}\int_{-\infty}^{\infty} \frac{n_{b,0}}{v_b}d\omega.
\end{align}
This is again a divergent quantity {at a fixed $z$.} However, important information about the phonon at $(\Omega, q)$ can be extracted by looking at the spectral density within a few $\Gamma$ around $\omega=0$ (since we have factored out the rapid oscillations at the carrier frequency $\Omega$). If we rewrite Eqn. \ref{eq:AF} in momentum space by using the relation $\omega=v_b k$ and change the variable of integration to $k$, Eqn. \ref{eq:AF} becomes exactly the same as Eqn. \ref{eq:AK}. Here linear dispersion is assumed. In more realistic situations, proper frequency cutoff can be set according to the actual dispersion relation of the system. A linear dispersion relation is a reasonable analytical assumption since {useful} information about the phonon mode is contained within a few $\Gamma$ around $\omega=0$.

As we will see in the following section, coupling between the acoustic field and optical fields modifies the acoustic mode at frequency $\Omega$ (or $\omega=0$ in the spectrum). However, due to the coupling between the acoustic modes, as the coupling increases the effect can spread out to other modes.

\section{Solution with non-depleted optical pumping}
Having discussed the behavior of the acoustic field without coupling, we are now in position to solve Eqns. \ref{eq:3WQ}. Assuming a non-depleted constant pump $a_1$, and defining the coupling constant $g=\beta a_1$, Eqns. \ref{eq:3WQ} can be simplified to
\begin{subequations}\label{eq:3WNDP}
	\begin{align}
	&\pd{a_2}{t}+v\pd{a_2}{z}
		=-\gamma a_2-igb, \label{eq:3WNDPa} \\
	&\pd{b}{t}+v_b\pd{b}{z}
		=-\Gamma b-ig^*a_2+\xi,
		\label{eq:3WNDPb}
	\end{align}
\end{subequations}
where we have changed $\gamma_2 \rightarrow \gamma$ and $v_2 \rightarrow v$ for the loss and the group velocity of the anti-Stokes field respectively since there is only one varying optical field under consideration. {We assume that at the beginning of the waveguide $(z=0)$ there is no anti-Stokes field present $a_2(0,t)=0$ and the noise-determined acoustic field $b(0,t)$ has a power spectral density given by Eqn.} \ref{eq:SbNCInf}.  In frequency space we have,
\begin{subequations}\label{eq:3WNDPFT}
	\begin{align}
	&-i\omega \tilde{a}_2+v\pd{\tilde{a}_2}{z}
		=-\gamma \tilde{a}_2-ig\tilde{b},
		\\
	&-i\omega \tilde{b}+v_b\pd{\tilde{b}}{z}
		=-\Gamma \tilde{b}-ig^*\tilde{a}_2+\tilde{\xi},
		\label{eq:3WNDPFTb}
	\end{align}
\end{subequations}
where the ``tilded'' variables are functions of both $z$ and $\omega$. Now we define the Laplace transform $\mathcal{L}\{ \}$ of these variables by
\begin{align}
A(s,\omega)\equiv\mathcal{L}\{\tilde{a}(z,\omega)\}=\int_{0}^{\infty}dz e^{-sz}\tilde{a}(z,\omega),
\end{align}
where we use capital letters to denote the Laplace transformed variables. Eqns. \ref{eq:3WNDPFT} become
\begin{subequations}\label{eq:3WNDPLT}
	\begin{align}
	&-i\omega A_2+vsA_2
		=-\gamma A_2-igB,
		\\
	&-i\omega B+v_bsB-v_b \tilde{b}(0,\omega)
		=-\Gamma B-ig^*A_2+\Xi,
		\label{eq:3WNDPLTb}
	\end{align}
\end{subequations}
where we have used initial conditions {as discussed above}. The solution of Eqs. \ref{eq:3WNDPLT} is
\begin{subequations}\label{eq:LTSol}
	\begin{align}
	&B=\frac{v_b \tilde{b}(0,\omega)+\Xi}
	{\Gamma-i\omega+v_bs+\frac{|g|^2}{\gamma-i\omega+vs}},\label{eq:LTSolb}
		\\
	&A_2=\frac{-igv_b \tilde{b}(0,\omega)-ig\Xi}
	{(\Gamma-i\omega+v_bs)(\gamma-i\omega+vs)+|g|^2}.
	\label{eq:LTSola}
	\end{align}
\end{subequations}
For mathematical convenience we define two complex functions
\begin{subequations}\label{eq:ILT}
	\begin{align}
	&f(z,\omega)=\mathcal{L}^{-1}\left \{\frac{1}{\Gamma-i\omega+v_bs+\frac{|g|^2}{\gamma-i\omega+vs}} \right \},
		\\
	&h(z,\omega)=\mathcal{L}^{-1}\left \{\frac{1}{(\Gamma-i\omega+v_bs)(\gamma-i\omega+vs)+|g|^2} \right \},
	\end{align}
\end{subequations}
where $\mathcal{L}^{-1}\{\}$ denotes the inverse Laplace transform. We can then formally solve for $\tilde{b}(z,\omega)$ and $\tilde{a}_2(z,\omega)$ by taking the inverse Laplace transform of Eqs. \ref{eq:LTSol} as follows:
\begin{subequations}\label{eq:FTSol}
	\begin{align}
	&\tilde{b}(z,\omega)=v_b \tilde{b}(0,\omega)f(z,\omega)+\int_{0}^{z}dz' f(z-z',\omega)\tilde{\xi}(z',\omega),
		\\
	&\tilde{a}_2(z,\omega)=-igv_b \tilde{b}(0,\omega)h(z,\omega)-ig \int_{0}^{z}dz' h(z-z',\omega)\tilde{\xi}(z',\omega).
	\end{align}
\end{subequations}
We can now evaluate $\langle \tilde{b}^\dagger(z,\omega)\tilde{b}(z,\omega') \rangle$ and $\langle \tilde{a}_2^\dagger(z,\omega)\tilde{a}_2(z,\omega') \rangle$ using the properties of the noise source in Eqns. \ref{eq:NSM} and \ref{eq:NS}. Again assuming that the correlations $\langle \tilde{b}^\dagger(0,\omega)\tilde{\xi}(z',\omega') \rangle$ and $\langle \tilde{\xi}^\dagger(0,\omega)\tilde{b}(z',\omega') \rangle$ vanish, we find the spectral density of the fields as follows
\begin{subequations}\label{eq:SDSol}
	\begin{align}
	&S_b(z,\omega)=v_b^2|f(z,\omega)|^2S_b(0,\omega)+2\Gamma n_{b,0}\int_{0}^{z}dz' |f(z-z',\omega)|^2,
		\\
	&S_{a_2}(z,\omega)=v_b^2|g|^2 |h(z,\omega)|^2S_b(0,\omega)+2\Gamma n_{b,0} |g|^2\int_{0}^{z}dz' |h(z-z',\omega)|^2,
	\end{align}
\end{subequations}
where $S_b(0,\omega)=n_{b,0}/v_b$.

\vspace{8pt}
The integrals in Eqns. \ref{eq:SDSol} can be integrated analytically, however, the closed form expression of is unwieldy and the complete expression provides little physical insight. An important parameter that characterizes the solution is given by $\sqrt{(\gamma/v-\Gamma/v_b)^2-4|g|^2/vv_b}$ which appears in exponential terms inside both $S_b$ and $S_{a_2}$. If we rewrite this factor in terms of spatial loss rates $\bar{\gamma}\equiv\gamma/v$, $\bar{\Gamma}\equiv\Gamma/v_b$ and define the spatial coupling rate $\bar{g}\equiv g/\sqrt{vv_b}$, the expression becomes $\sqrt{(\bar{\gamma}-\bar{\Gamma})^2-4|\bar{g}|^2}$. Expressions of this form are commonly found in optomechanical systems \cite{tomes2011quantum, rakich2016quantum}. {Analogously, the coupling strength in this system is characterized by the discriminant}
\begin{align}
\Delta=(\bar{\gamma}-\bar{\Gamma})^2-4|\bar{g}|^2.
\end{align}
For low coupling $\Delta>0$, the opto-acoustic interaction adds an effective spatial loss to the acoustic and optical fields. For high coupling $\Delta<0$, {in addition to the effective spatial loss} the solution becomes oscillatory within the decay length $1/(\bar{\gamma}+\bar{\Gamma})$, having a spatial period of $2\pi/\sqrt{|\Delta|}$. Note that for the solution to be valid the coupling strength should be bounded by the relation $\sqrt{|\Delta|}\ll q, k_2$. This is because if the coupling $\bar{g}$ approached a value such that $\sqrt{|\Delta|}\gtrsim q$ or $\sqrt{|\Delta|}\gtrsim k_2$, there would be rapid oscillations within one wavelength of the carrier field and the slowly varying approximation would break down.

\begin{figure}[th!]
\begin{center}
\includegraphics[trim=0cm 1.3cm 0cm 2cm, clip, width=1\textwidth]{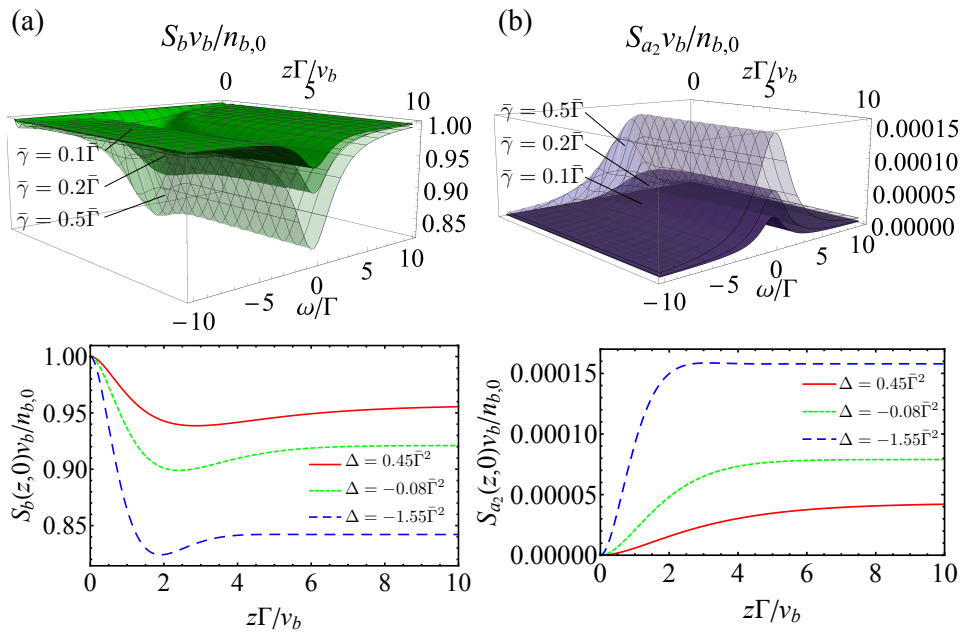}
\caption{Spectra of (a) phonon density and (b) anti-Stokes photon density as functions of $z$ and $\omega$ for cooling of high group velocity phonons. We set $\gamma=1000\Gamma$ and $g=30\Gamma$. The group velocity of the acoustic mode is set to $v_b=10^{-4}v$, $v_b=2\times 10^{-4}v$, and $v_b=5\times 10^{-4}v$ corresponding to spatial loss rates of $\bar{\gamma}=0.1\bar{\Gamma}$, $\bar{\gamma}=0.2\bar{\Gamma}$, and $\bar{\gamma}=0.5\bar{\Gamma}$. The corresponding discriminants are $\Delta= 0.45 \bar{\Gamma}^2$, $\Delta= -0.08 \bar{\Gamma}^2$, and $\Delta= -1.55 \bar{\Gamma}^2$, respectively. The bottom panel of the figure shows the evolution of the spectrum at center frequency $\omega=0$. With larger group velocity $v_b$, the acoustic spectrum exhibits oscillatory behavior near the beginning of the waveguide corresponding to a negative value of $\Delta$.
		}
\label{fig:SpecHVG}
\end{center}
\end{figure}

\vspace{8pt}
The behavior of the spectral density given by Eqns. \ref{eq:SDSol} depends strongly the group velocity $v_b$ of the acoustic field. In Fig. \ref{fig:SpecHVG} we plot the spectral density of the acoustic and optical fields as functions of $z$ and $\omega$ with $\gamma=1000\Gamma$, and $g=30\Gamma$. The values of temporal loss rate and coupling used here are typically found in optomechanical systems and resonators \cite{agarwal2013multimode}. The $v_b$ of the phonon mode is set to {correspond to high group velocity phonons}. Since the coupling rate is fixed at $g=30\Gamma$, the effective spatial coupling is given by $\bar{g}=30\bar{\Gamma}\sqrt{v_b/v}$. We see that the effect of increasing $v_b$ leads to an increase in the spatial coupling $\bar{g}$ relative to the spatial acoustic loss $\bar{\Gamma}$. For $v_b=5\times 10^{-4}v$, the spectral density of the phonon reaches 0.85 at $\omega=0$ and $z\rightarrow \infty$. The anti-Stokes spectral density in this case builds up rapidly within a few {acoustic }decay lengths and reaches its steady value.

\begin{figure}[th!]
\begin{center}
\includegraphics[trim=0cm 1.8cm 0cm 2cm, clip, width=1\textwidth]{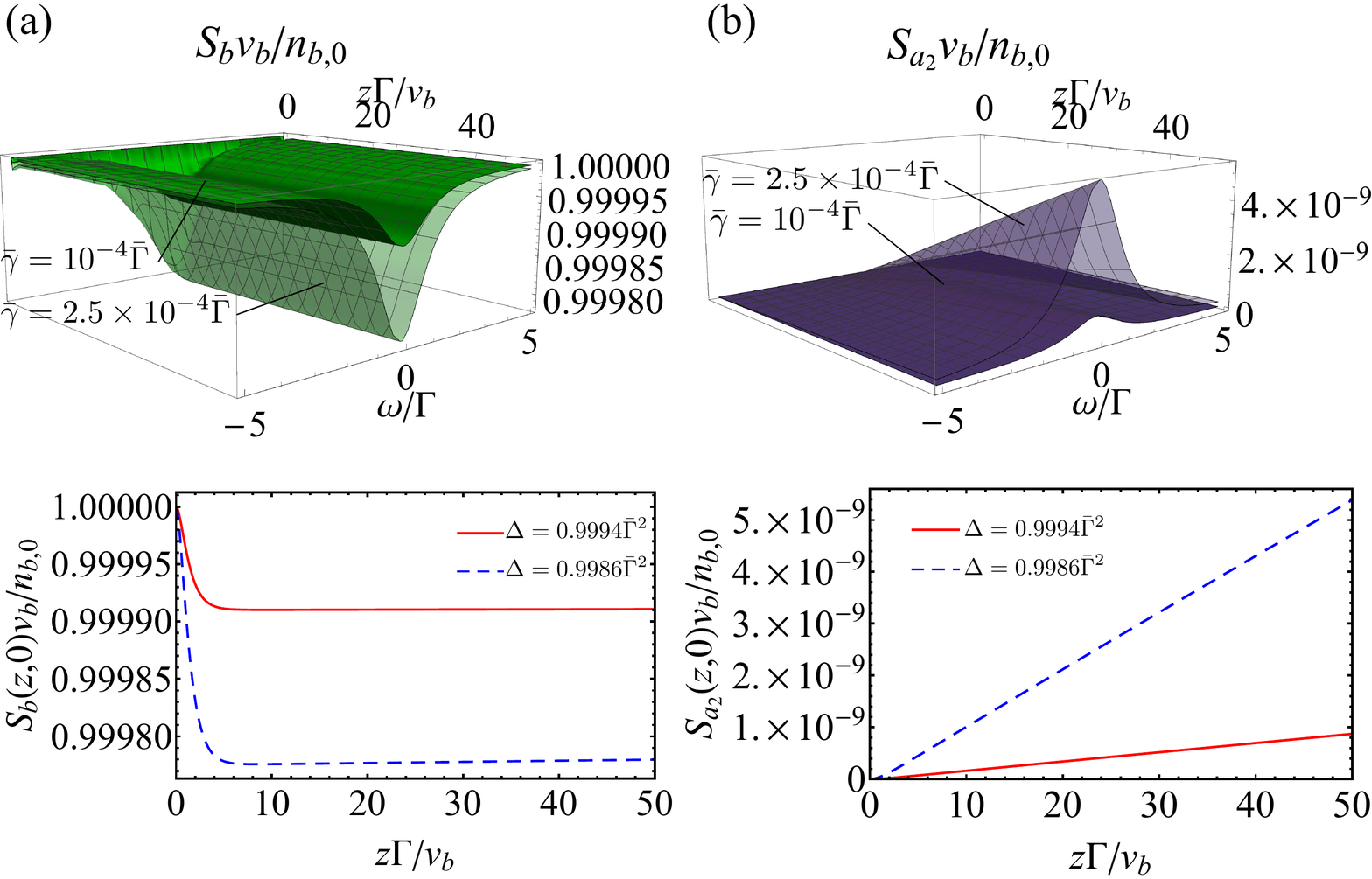}
\caption{Spectra of (a) phonon density and (b) anti-Stokes photon density as functions of $z$ and $\omega$ for low group velocity phonons. We set $\gamma=1000\Gamma$ and $g=30\Gamma$. The group velocity of the acoustic mode is set to $v_b= 10^{-7}v$ and $v_b=2.5\times 10^{-7}v$ corresponding to spatial loss rates of $\bar{\gamma}=10^{-4}\bar{\Gamma}$ and $\bar{\gamma}=2.5 \times 10^{-4}\bar{\Gamma}$. The corresponding discriminants are $\Delta= 0.9994 \bar{\Gamma}^2$ and $\Delta= 0.9986 \bar{\Gamma}^2$, respectively.  The bottom panel of the figure shows the evolution of the spectrum at center frequency $\omega=0$. For low group velocity phonons the effective coupling per unit length $\bar{g}$ is almost zero.
		}
\label{fig:SpecLVG}
\end{center}
\end{figure}

Fig. \ref{fig:SpecLVG} shows the spectral density of the acoustic and optical fields for low group velocity phonons with all other parameters the same as Fig. \ref{fig:SpecHVG}. For these phonon velocities the interaction between the fields is almost negligible. The acoustic spectral density is barely perturbed at the center frequency $\omega=0$, and the anti-Stokes light builds up almost linearly.

\vspace{10pt}
The degree of phonon cooling can be extracted if we write the spectral density in a form similar to Eqn. \ref{eq:SbNCInf} as in $S_b(z,\omega)\equiv n_b(z,\omega)/v_b$, where $n_b(z,\omega)$ can be understood as effective phonon spectral occupation at position $z$. Since the energy per quanta $\hbar\Omega$ is much lesser than $k_B T$ we can further approximate $n_b(z,\omega)\approx k_B T_{\text{eff}}(z,\omega)/\hbar\Omega$, where $T_{\text{eff}}(z,\omega)$ is an effective local temperature for the mode. The ratio of the spectral density $S_b(z,\omega)$ to $S_b(0,\omega)$ is then simply the cooling ratio
\begin{align}
\frac{S_b(z,\omega)}{S_b(0,\omega)}=\frac{S_b(z,\omega)v_b}{n_{b,0}}=\frac{n_b(z,\omega)}{n_{b,0}}\approx \frac{T_{\text{eff}}(z,\omega)}{T_0},
\end{align}
where $T_0$ is the temperature of the bath. The degree of cooling over certain bandwidth $\Delta \omega$ around the central phonon frequency is then given by
\begin{align}
\frac{\displaystyle \int_{-\Delta \omega/2}^{\Delta \omega/2}S_b(z,\omega)d\omega}{\displaystyle\int_{-\Delta \omega/2}^{\Delta \omega/2}S_b(0,\omega)d\omega}=\frac{v_b \displaystyle \int_{-\Delta \omega/2}^{\Delta \omega/2}S_b(z,\omega)d\omega }{n_{b,0}\Delta \omega}.
\end{align}

\begin{figure}[th!]
\begin{center}
\includegraphics[trim=0.6cm 4.1cm 0.5cm 4.3cm, clip, width=1\textwidth]{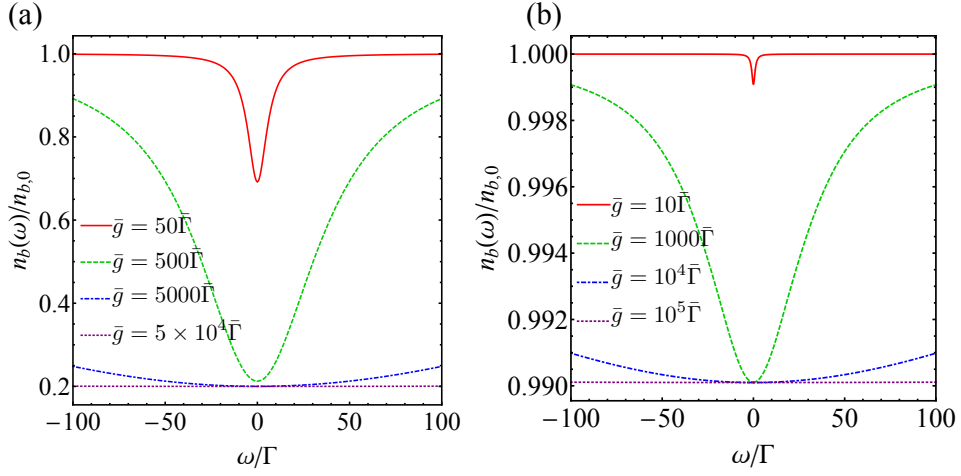}
\caption{Observing the limit of spectral cooling ratio at large $z$. (a) Spectral cooling ratio $n_b(\omega)/n_{b,0}$ at large distance $z\Gamma/v_b=5$ for spacial loss rate $\bar{\gamma}=4\bar{\Gamma}$. The maximum ratio (Eqn. \ref{eq:Lim}) in this case is 0.2. (b) Spectral cooling ratio $n_b(\omega)/n_{b,0}$ at large distance $z\Gamma/v_b=100$ with lower spatial loss rate $\bar{\gamma}=0.01\bar{\Gamma}$. The maximum ratio (Eqn. \ref{eq:Lim}) in this case is 0.99.
		}
\label{fig:Lim}
\end{center}
\end{figure}

\vspace{8pt}
In Fig. \ref{fig:Lim} we plot the acoustic spectral density at a fixed position with increasing coupling $\bar{g}$. As the coupling increases, the dip in the acoustic spectrum corresponding to the reduced phonon occupation is observed to  broaden and the spectrum at $\omega=0$ reaches a limiting value. In fact, this limit takes the same form as the limit for cooling in resonators \cite{tomes2011quantum, aspelmeyer2014cavity} but with the temporal loss rates replaced by their corresponding spatial loss rates:
\begin{align}\label{eq:Lim}
\lim_{g\rightarrow\infty}\frac{n_{b}(z\rightarrow \infty, \omega)}{n_{b,0}}=\frac{\bar{\Gamma}}{\bar{\Gamma}+\bar{\gamma}}.
\end{align}
Physically this means that the coupling between the fields introduces additional effective damping per unit length to the acoustic field, but the largest possible damping is limited by the optical loss rate.

\begin{figure}[th!]
\begin{center}
\includegraphics[trim=0cm 4.3cm 0.2cm 4.4cm, clip, width=1\textwidth]{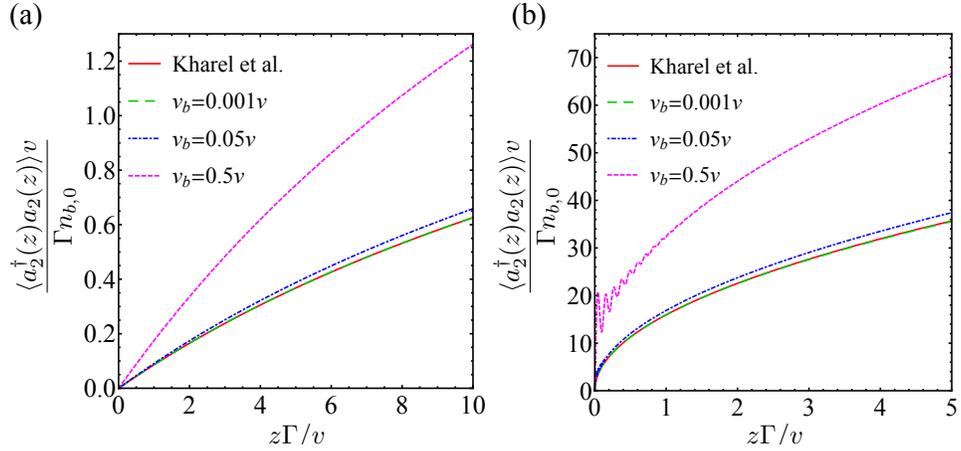}
\caption{Anti-Stokes photon density as a function of position. Here optical loss $\gamma$ is set to zero, and no cooling of the phonon will occur at large $z$. (a) Low coupling case $g=0.3\Gamma$ and (b) moderate coupling case $g=20\Gamma$. In both cases, as $v_b$ decreases, the solution for anti-Stokes photon density approaches Kharel's solution \cite{kharel2016noise}.
		}
\label{fig:ASC}
\end{center}
\end{figure}

\vspace{12pt}
We conclude this section with a comparison between our solution and that of a recent study on noise in Brillouin systems \cite{kharel2016noise}, which only solves for low group velocity phonons and in which cooling is not discussed. The optical loss was set to zero in the calculation of \cite{kharel2016noise} which also means that no phonon cooling can occur (Eqn. \ref{eq:Lim}). For finite coupling $g$, we can integrate the anti-Stokes spectrum to obtain the anti-Stokes photon number {per unit length} $\langle a_2^\dagger(z,t) a_2(z,t) \rangle$ and compare with the solution given in \cite{kharel2016noise}, which is
\begin{align}\label{eq:KS}
\langle a_2^\dagger(z,t) a_2(z,t) \rangle =\frac{n_{b,0} |g|^2z}{v^2}\exp\left (-\frac{|g|^2z}{v\Gamma}\right )\left [I_0 \left (\frac{|g|^2z}{v\Gamma}\right ) +I_1 \left (\frac{|g|^2z}{v\Gamma}\right )\right ].
\end{align}
In Eqn. \ref{eq:KS}, $I_0$ and $I_1$ are the modified Bessel functions of the first kind. Fig. \ref{fig:ASC} compares the solutions for low and moderate coupling. In both cases as we decrease the group velocity of the phonons, our solution approaches that of \cite{kharel2016noise}. Note that we use $v$ instead $v_b$ to normalize the variables in Fig. \ref{fig:ASC} since $v_b$ is set to zero in \cite{kharel2016noise}.

\FloatBarrier

\section{Practical considerations for {observing Brillouin} cooling in linear waveguides}

Now that we understand the general behavior of the solution, we can relate the calculation to available experimental data.  To relate the coupling coefficient $g=\beta a_1$ to commonly measured experimental quantities, we first note that Eqns. \ref{eq:3WQ} can also describe a Stokes process if we let $a_2$ be the pump field and $a_1$ be the Stokes field.

In a standard SBS experiment the strength of the interaction is characterized by the SBS gain $G$ (W$^{-1}$m$^{-1}$), which is given by the relation
\begin{align}\label{eq:Gain}
\pd{P_1}{z}=GP_2P_1,
\end{align}
where we use $P_1$ and $P_2$ to denote the Stokes power and pump power respectively to be consistent with Eqns. \ref{eq:3WQ}. For typical experimental SBS setups \cite{shin2013tailorable,van2015interaction,van2015net} the equations for the Stokes and the acoustic fields under steady condition with strong pump are given by
\begin{subequations}
	\begin{align}
	&v\pd{a_1}{z}
		=-ig^*b^\dagger, \\
	&v_b\pd{b}{z}
		=-\Gamma b-ig^*a_1^\dagger,
	\end{align}
\end{subequations}
where the optical loss is usually neglected \cite{kharel2016noise,PhysRevA.92.013836,wolff2015acoustic} and we have used $v_1\approx v_2\approx v$ for the optical fields. {Note that since we are dealing with Stokes process here, we have $g=\beta a_2^\dagger$ in the above equations.} For phonon modes with large spatial loss, which is typically the case in the experiments, the spatial variation of $b$ is treated as negligible. We can immediately solve $b=-ig^*a_1^\dagger/\Gamma$. Substituting this into the Stokes equation gives
\begin{align}
\pd{a_1}{z}=\frac{|g|^2}{v\Gamma}a_1.
\end{align}
Comparing this with the SBS gain relation Eqn. \ref{eq:Gain} we find
\begin{align}\label{eq:CC}
g=\sqrt{\frac{GP_2v\Gamma}{2}},
\end{align}
which allows us to relate the coupling coefficient $g$ to SBS gain and pump power $P_2$. For Brillouin scattering since $\Omega\ll\omega_1,\omega_2$, this gain $G$ is symmetric for Stokes and anti-Stokes scattering. {It is worth noting that the coupling coefficient $g$ (through $\beta$) is given by the overlap integral between the optical and acoustic modes and is thus independent of the acoustic loss $\Gamma$. The relation in Eqn.} \ref{eq:CC} {is a good approximation for relating the coupling coefficient to the SBS gain in typical SBS experiments. It may not be a good approximation, however, when the phonon has large group velocity or extremely low loss. Presently, this relation allows us to estimate how strong the coupling could be in recently state-of-the-art SBS systems.}

\vspace{14pt}
We can now use known experimental results to evaluate how much cooling could potentially be achieved in a waveguide. In the following discussion we fix the pump wavelength at 1550 nm and focus only on silicon photonic waveguides \cite{shin2013tailorable,van2015interaction,van2015net}. {The group velocity of the optical field in these waveguides is approximately $v\approx c/n\approx 8.62\times 10^7$ m/s for low dispersion modes, but is certainly dependent on the specific optical mode}. Since the optical decay length in photonic waveguides is typically of the order of cm, we fix the optical loss at $\bar{\gamma}$ at 3.456 m$^{-1}$ corresponding to a good value of 0.3 dB/cm reported in \cite{vlasov2004losses}.

The acoustic loss can be obtained from the $Q$-factor of the acoustic modes using the expression $\bar{\Gamma}=\Omega/2v_bQ$. The theoretical limits on the acoustic $Q$-factor in silicon have been suggested previously in \cite{tabrizian2009effect}. Using these data, for a 500 MHz phonon the $Q$-factor for silicon can theoretically approach 50000, giving an acoustic loss of 3.7 m$^{-1}$ for phonons traveling at the speed of sound $v_b= 8500$ m/s. However, the $Q$-factors of the phonon modes in recent experiments are not so high, and are typically of the order of a few hundreds \cite{shin2013tailorable,van2015interaction,van2015net}. Since these prior studies are for low group velocity phonons $\sim 1$ m/s, it will lead to acoustic loss of the order $\sim10^{6}$ m$^{-1}$ for phonons in the 500 MHz range. 
As we show above, higher group velocity phonon modes are more practical for cooling.
For the sake of calculation we fix the velocity at $v_b\approx 8500$ m/s \cite{tabrizian2009effect} {corresponding to traveling acoustic phonons in silicon} but assume a low $Q$-factor ($Q\sim 100$), giving us an acoustic loss of 1848 m$^{-1}$.

Recently reported SBS gain in silicon waveguides ranges from 1000 W$^{-1}$m$^{-1}$ \cite{shin2013tailorable,van2015interaction} to 10$^{4}$ W$^{-1}$m$^{-1}$ \cite{van2015net}. Gain as high as 4 $\times$ 10$^{6}$ W$^{-1}$m$^{-1}$ was also reported for acousto-optic scattering from a 6 MHz phonon mode in a specialized optical fiber \cite{butsch2014cw}.
the above parameters now allow evaluation of $g$ through Eqn. \ref{eq:CC}.

\begin{figure}[th!]
\begin{center}
\includegraphics[trim=0cm 1.2cm 0cm 2cm, clip, width=1\textwidth]{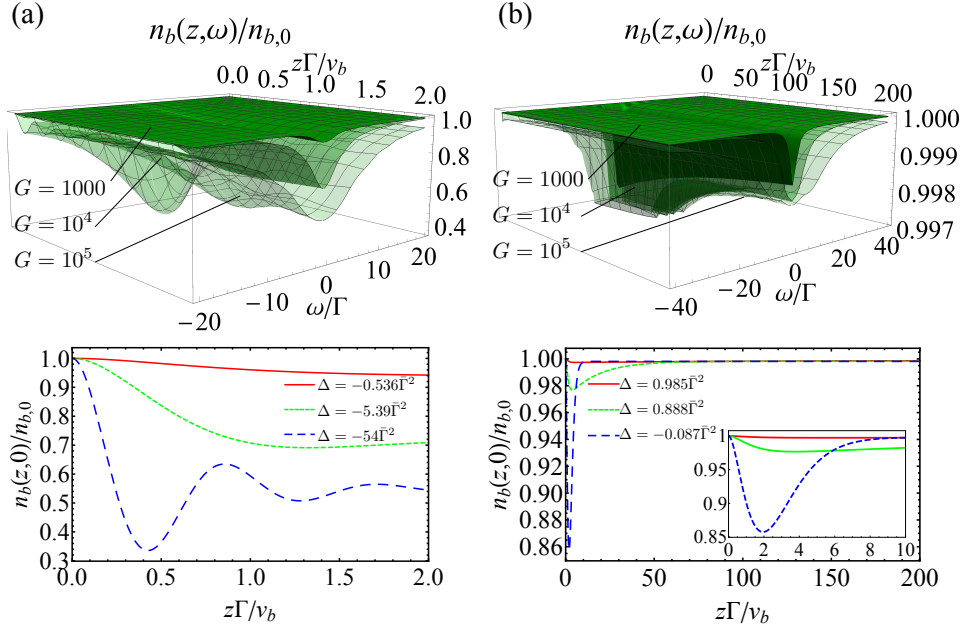}
\caption{Spectral cooling ratio of 500 MHz phonons for (a) a high-$Q$ acoustic mode $Q=50000$ ($\bar{\Gamma}=3.7$ m$^{-1}$), with 1 mW pump power, and (b) a low-$Q$ acoustic mode $Q=100$ ($\bar{\Gamma}=1848$ m$^{-1}$), with 10 mW pump power. In both cases optical parameters are $\bar{\gamma}=3.456$ m$^{-1}$ and $v=8.62\times 10^7$ m/s, while phonon group velocity is set to a high number $v_b=8500$ m/s.  The bottom panel of the figure shows the evolution of the spectrum at center frequency $\omega=0$. The inset in (b) shows that the cooling ratio for high SBS gain  $G=10^5$ W$^{-1}$m$^{-1}$ can reach far below the long range limit in the beginning portion of the waveguide.
		}
\label{fig:GP}
\end{center}
\end{figure}

Using the parameters described above, we can now plot (in Fig. \ref{fig:GP}) the spectral cooling ratio for acoustic modes for both $Q=50000$ and $Q=100$ at 500 MHz with various SBS gain $G$ and pump power $P_1$. For the high-$Q$ mode (Fig. \ref{fig:GP}a) with large gain, cooling at the central frequency can reach the saturation limit of 0.517 within a few decay lengths with just 1 mW pump. The spectrum in this strong coupling regime also exhibits oscillatory behavior as discussed previously. For the low $Q$ mode (Fig. \ref{fig:GP}b), the cooling limit 0.9981 is reached after a few hundred decay lengths with a larger 10 mW pump. Note that at the entrance region of the waveguide the cooling ratio can momentarily dip far below the limit given by Eqn. \ref{eq:Lim}, but it will return to this limit at larger $z$. These results show that to observe appreciable cooling in an experiment, a high-$Q$ phonon with large group velocity is required. Similar calculation can be carried out for low group velocity phonons ($v_b\sim$1 m/s), however, no appreciable cooling can be achieved even with modes with $Q$ as high as 50000.

\FloatBarrier

\section{Conclusions}

We have investigated the anti-Stokes Brillouin scattering process in linear waveguides for the possibility of achieving phonon annihilation and cooling. We show that the degree of cooling depends strongly on the acoustic and optical spatial loss rates. The results can be categorized into two regimes that depend on the strength of the acousto-optical interaction: strong coupling ($\Delta>0$) and low coupling ($\Delta<0$). In the first case the spectrum of the fields shows oscillatory behavior before reaching steady-state. In the second case, the spectrum exhibits no oscillatory behavior and only reaches its steady-state after longer propagation distances. The limit on the cooling ratio in waveguides takes the same form as in optomechanical cooling in resonators, as expected, with the temporal variables replaced by corresponding spatial loss variables. The effect of higher phonon group velocity is to not only increase the spatial coupling strength but also give an improved cooling ratio limit. 

Our analysis indicates that for potential cooling experiments, traveling phonons with high group velocity are more favorable, and the design of waveguides should be focused on improving the quality factor of such high group velocity phonon modes. Additionally, enhancement of SBS gain is also highly desirable for cooling \cite{dostart2015giant,rakich2012giant}. With a combination of high-$Q$, high group velocity phonons, and gain of the order of 10$^{5}$ W$^{-1}$m$^{-1}$, appreciable cooling may be observed with a pump power of a few mW.

A key difference between cavity optomechanical cooling and Brillouin cooling is that the latter annihilates traveling phonons, whose higher frequency cousins are also responsible for heat transport in micro- and nano-scale devices. The possibility of accessing Brillouin cooling in linear systems thus opens up a new approach to laser cooling, and may in the future impact phonon management in photonic circuits and quantum information transport via phonons. 

\section*{Acknowledgement}
The authors would like to thank Prof. P. Scott Carney for stimulating scientific discussions and for encouraging us along in this work. This work was funded by a US Army Research Office grant (W911NF-15-1-0588).

\bibliographystyle{ieeetr}
\bibliography{SBSWG}
\end{document}